\documentclass[twocolumn,english,amssymb,aps,superscriptaddress,showpacs,amsmath,showkeys,floatfix,pra]{revtex4-1}
\usepackage[T1]{fontenc}
\usepackage[latin9]{inputenc}
\usepackage{geometry}
\usepackage[active]{srcltx}
\usepackage{amsmath}
\usepackage{graphicx}
\usepackage{esint}
\usepackage{epsfig}
\usepackage{epstopdf}
\usepackage{physics}
\usepackage{color}
\usepackage{natbib}
\usepackage{upgreek}
\usepackage{mathtools}
\usepackage{soul}
\usepackage{lipsum}
\usepackage{enumitem}
\usepackage{amsfonts}
\usepackage{eurosym}
\usepackage{braket}
\usepackage{setspace}
\newcommand {\rttensor}[1]{\overline{\overline{#1}}}
\newcommand\identity{1\kern-0.25em\text{l}}
\newcommand{\bs}{\boldsymbol}
\usepackage[dvipsnames]{xcolor}
\usepackage{bm}
\usepackage{ulem}
\usepackage{color}
\definecolor{blue}{rgb}{0,0,1}
\definecolor{green}{rgb}{0,1,0}
\definecolor{purple}{rgb}{0.5,0,1}
\makeatletter
\usepackage{babel}
\makeatother

\begin{document}

\title{Anomalous high-density spin noise in a strongly interacting atomic vapor}

\author{J. Delpy}
\affiliation{Universit\'e Paris-Saclay, CNRS, Ecole Normale Sup\'erieure Paris-Saclay, CentraleSup\'elec, LuMIn, Orsay, France}
\author{N. Fayard}
\affiliation{Universit\'e Paris-Saclay, CNRS, Ecole Normale Sup\'erieure Paris-Saclay, CentraleSup\'elec, LuMIn, Orsay, France}
\author{F. Bretenaker}
\affiliation{Universit\'e Paris-Saclay, CNRS, Ecole Normale Sup\'erieure Paris-Saclay, CentraleSup\'elec, LuMIn, Orsay, France}
\author{F. Goldfarb}
\affiliation{Universit\'e Paris-Saclay, CNRS, Ecole Normale Sup\'erieure Paris-Saclay, CentraleSup\'elec, LuMIn, Orsay, France}

\begin{abstract}

 Spin noise spectroscopy (SNS) has become a mainstream approach to probe the dynamics of a spin ensemble in and out of equilibrium. Current models describing spin noise in interacting samples are based on an effective single particle dynamics in a bath. Here, we report the observation of a strong interaction regime which significantly affects the spin dynamics. Performing SNS in a dense Rubidium vapor, we observe anomalous distortions of the usual spin noise spectra, which we attribute to resonant dipole-dipole interaction within the ensemble. As the density of the vapor increases, we observe a dramatic broadening of the usual resonances and the emergence of an unexpected extra low-frequency noise component. We use a simple microscopic two-body numerical model to reproduce and discuss these observations. Our results suggests that the spectra cannot be described by usual models of single-atom dynamics and arise from the evolution of interacting pair of atoms. This work opens the way to the study of many-body spin noise or higher order correlators in atomic vapors using SNS.

\end{abstract}

\maketitle
\section{Introduction}

The optical measurement of spontaneous angular momentum fluctuations in particle ensembles, called spin noise spectroscopy (SNS) \cite{aleksandrov_magnetic_1981, crooker_spectroscopy_2004}, is a powerful approach to the characterization of spin dynamics. First introduced in dilute atomic vapors \cite{ katsoprinakis_measurement_2007, mihaila_quantitative_2006, fomin_spin-alignment_2020, liu_birefringence_2022, liu_spin-noise_2023}, it was extended to condensed phases such as semiconductors \cite{hubner_rise_2014}, quantum wells \cite{poltavtsev_spin_2014} or quantum dots \cite{crooker_spin_2010, glasenapp_spin_2016, gundin_spin_2024}. This method quickly gave new insights into the relaxation and decoherence channels appearing in such systems. In atomic ensembles, SNS reveals ground-state energy structures both in non-invasive \cite{crooker_spectroscopy_2004,zapasskii_optical_2013, swar_detection_2021} and invasive \cite{glasenapp_spin_2014,chalupczak_near-resonance_2011,horn_spin-noise_2011,swar_measurements_2018} regimes. The latter has gained considerable interest in the last decade, since it allows to characterize the interaction of the system with its environment. Indeed, driving a system out of equilibrium allows to obtain information beyond the fluctuation-dissipation theorem \cite{sinitsyn_theory_2016, li_nonequilibrium_2013}. For instance, the coupling induced by a radiofrequency field between Zeeman sublevels was characterized using SNS \cite{glasenapp_spin_2014}. Noisy magnetic fields were also shown to give rise to non-thermal spin noise in weakly-pumped atomic ensemble \cite{delpy_creation_2023} and even non-Gaussian spin dynamics, as revealed by higher-order SNS \cite{li_higher-order_2016}.

However, the use of SNS to probe interactions \textit{within} an ensemble has been mainly restricted to an extensive characterizations of spin-exchange (SE) collisions in single-species systems \cite{katsoprinakis_measurement_2007, mouloudakis_quantum_2019, mouloudakis_effects_2022} and two-species systems \cite{mouloudakis_interspecies_2023, roy_probing_2015}. Spin-exchange originates from the Pauli exclusion principle and is thus a short-range interaction, with a cross-section $\sigma \propto a_0^2$ where $a_0$ is the Bohr radius. As a consequence, such an impact-like interaction is so short-lived that the spin dynamics remains essentially that of independent particles. Therefore, spin noise spectra can be inferred from mean-field descriptions of the evolution of single-particle operators \cite{happer_effect_1977, grossetete1, grossetete2} in which the SE simply materializes as a change in the spin decoherence rate \cite{katsoprinakis_measurement_2007}. Only in the spin-exchange relaxation free (SERF) conditions has this picture recently been proven incomplete. In this regime, arising spin correlations lead to changes in the spin noise (SN) lineshapes, which can be described by mean-field hyperfine spin equations of evolution \cite{mouloudakis_effects_2022, mouloudakis_interspecies_2023, mouloudakis_anomalous_2024}. Moreover, entanglement induced by SE collisions was studied theoretically \cite{mouloudakis_spin-exchange_2021} but lacks experimental evidence. Therefore, the capability of SNS to probe collective effects beyond the single-particle description needs to be further investigated in the search for many-body regimes or higher-order spin correlations in vapors.

In this article, we investigate the spin noise in a system in which stronger, long-range interaction are expected to occur. We perform SNS in a high-density Rubidium cell close to the $\mathrm{D}_2$ optical transition, where resonant dipole-dipole coupling between different atoms can kick in. This long-range interaction, never observed until now in SNS, is well known to produce a collective optical response in vapors at high densities \cite{gross_superradiance_1982, rohlsberger_collective_2010, Keaveney_2012, peyrot_collective_2018}. We thus study the potential manifestation of such a collective regime in the spin noise spectrum of the vapor. A following fundamental question we consider is whether these collective effects simply lead to a renormalization of a single-particle spectrum, or if they result in new spectral features. This paper investigates these points in the following way: section II describes the experimental SNS apparatus and first reports unexpected, density-dependent changes at high densities. Section III introduces a microscopic model to simulate the spin fluctuations of a system of two atoms coupled by resonant dipole-dipole interactions, including atomic motion. Finally, section IV compares simulations and experiments and discusses how the anomalous changes in the experimental spin noise spectra can emerge from the coupling of the atoms. Finally, we propose a possible explanation to the low-frequency noise and discuss its implications in regard of a correlated spin noise regime.

\section{Experimental observations of collective spectra}

\begin{figure}
    \centering
    \includegraphics[width=\columnwidth]{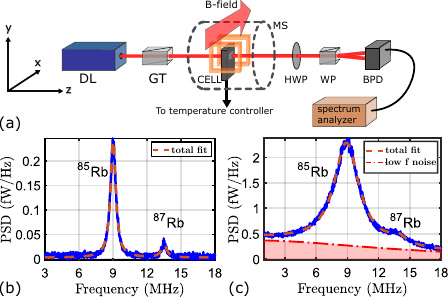}
    \caption{(a) Experimental setup used for SNS experiments. DL : diode laser, GT : Glan-Thompson polarizer, MS : magnetic shielding, HWP : half-wave plate, WP : Wollaston prism, BPD : balanced photodetection. (b,c) Spin noise spectra obtained for a cell temperature (b) $T=90^{\circ}$C and (c)  $T=175^{\circ}$C. The shaded area in (c) shows the low-frequency noise discussed in the text.}
    \label{figure1}
\end{figure}

A schematics of the apparatus is shown in figure \ref{figure1}\,(a). An external cavity diode laser is tuned at 780 nm with a power of a few mW to probe spin noise in the vicinity of the $\mathrm{D}_2$ line of Rubidium. We use a Glan-Thomson polarizer to ensure a good linear polarization of the beam. It is then focused into the vapor cell to reach a diameter of about $\phi \simeq 300$ \textmu m. A 1-mm- thick vapor cell is embedded in a metallic oven, the temperature of which is fine-tuned using a homemade temperature controller. The typical achievable temperatures range from ambient temperature to around 180$^{\circ}$C.
We probe the spin noise in the vapor by measuring the stochastic Faraday-like rotations of the probe polarization $\delta \theta(t) \propto \langle S_z \rangle (t)$ where $S_z$ is the projection of the collective spin on the laser propagation axis $z$ \cite{aleksandrov_magnetic_1981, crooker_spectroscopy_2004, mihaila_quantitative_2006}. A half-wave (HW) plate followed by a Wollaston prism (WP) splits the beam at the output of the cell in two well-balanced paths of orthogonal polarization, which are then sent on two photodiodes. The difference of the photocurrents is then proportional to the tiny Faraday rotation angle: $\Delta I = I_+ - I_- \propto \delta \theta (t)$. We feed this differential photocurrent in an electrical spectrum analyzer (ESA) to get the SN spectrum of the atomic vapor, i.e. the power spectral density PSD($S_z$). Following usual SNS setups \cite{aleksandrov_magnetic_1981, crooker_spectroscopy_2004}, a pair of coils is used to create a transverse magnetic field aligned along the $x$ axis, forcing the spontaneous spin coherences to oscillate at the Larmor frequency, typically a few MHz. Consequently, the spin noise resonances appear in a region of the spectrum which is free from low-frequency technical noises other than the laser shot noise and the ESA electronic noise background.

A typical example of SN spectrum obtained at relatively low density  $n \simeq 10^{12}\,\mathrm{at.cm}^{-3}$ (T = 90$^{\circ}$C) can be seen in Fig.\,\ref{figure1}\,(b). The probe power is 2\,mW and the magnetic field is $B\simeq 19\,\mathrm{G}$. In the following, we define the detuning as $\Delta = \omega_p - \omega_0$ with $\omega_p$ the probe laser frequency and $\omega_0$ the $F=3\rightarrow F'$ hyperfine transition frequency. This spectrum was obtained for $\Delta/2\pi = 1.5\,\mathrm{GHz}$. This corresponds to a probe laser tuned halfway between the $F=2\rightarrow F'$ and $F=3\rightarrow F'$ hyperfine transitions between the  $5^2S_{1/2}$ ground state and $5^2P_{3/2}$ excited state of $^{85}\mathrm{Rb}$, meaning that all ground-state hyperfine levels participe to the spin noise signal. This detuning is also much larger than the Doppler broadening of the ground-state hyperfine transitions and thus the absorption is negligible. The laser shot noise floor has been subtracted from the data. The spin noise spectrum then shows one SN resonance for each isotope, the width of which is mainly determined by the transit rate of the atoms through the laser beam in ballistic regime: $\gamma_t/2\pi \simeq \bar{v}/(2\pi\Phi) = 250\,\mathrm{kHz}$. Here $\bar{v} = \sqrt{3k_b T/m}$ is the average thermal velocity in the vapor. This simple spectrum is very well understood and arises from the stochastic precession of a single spin in a transverse magnetic field, under relaxation processes such as transit decoherence or SE collisions.

In order to probe higher density regimes, we perform SNS in a 1-mm-thick cell, thin enough to obtain low optical depth even close to resonance and thus reduce light absorption. Consequently, we are able to work at large atomic densities (up to around $4\times 10^{14} \,\mathrm{at.cm}^{-3}$ at $T=180^{\circ}\mathrm{C}$) while keeping the detuning as low as $\Delta/2\pi = 1.5\,\mathrm{GHz}$. This would not be possible using typical cm-long spectroscopy cells because of the complete extinction of the beam. At such densities in our experiment, absorption is not negligible but the output transmission (typically 25 \%) is still substantial. In this configuration, we report drastic changes in the spin noise lineshapes when the density of the vapor increases. A SN spectrum recorded in similar conditions to that of Fig.\,\ref{figure1}(b) except for a higher temperature of T=$ 175^{\circ}\mathrm{C}$ is shown in Fig.\,\ref{figure1}(c). It shows two important features: (i) the \textit{large broadening of both spin resonance lines}, up to widths of the order of $1\,\mathrm{MHz}$, and (ii) the emergence of a broad  \textit{low-frequency noise component} never-observed in SNS to the best of our knowledge. To characterize these unusual lineshapes, we fit each spectrum by a function consisting in the sum of three lines. The first one is a broad Lorentzian line centered on the zero frequency, which accounts for the extra low-frequency noise. The two other lines correspond to the spin resonance peaks of both $^{85}$Rb and $^{87}$Rb isotopes, and are given by:
\begin{equation}
    C_{zz}(\omega) = A\times\dfrac{\gamma(\gamma^2+\omega^2 +\omega_L^2) }{(\gamma^2 + \omega_L^2-\omega^2)^2 + 4\gamma^2\omega^2}.
    \label{equation1}
\end{equation}
These specific lineshapes arise from the Bloch equations describing the damped precessions of spins around a magnetic field aligned with the $x$ axis, under the action of a white-noise forcing term. The Bloch equations can be written under the matrix form
\begin{equation}
    \bm{\dot{S}} = R \bm{S} + G \bm{\eta},
    \label{equation2}
\end{equation}
with $\bm{S} = \begin{pmatrix} S_y \\ S_z \end{pmatrix}$, $R = \begin{pmatrix} -\gamma & \omega_L \\ -\omega_L & -\gamma \end{pmatrix}$ the deterministic drift matrix, and $\bm{\eta} = \begin{pmatrix} \eta_y \\ \eta_z \end{pmatrix}$ a stochastic variable with a zero mean and unit variance, accounting for the spin fluctuations. The matrix $G$ describes the strength of the spin noise, and we take $G \equiv \sqrt{A\gamma}\times \identity$, $A$ being a free parameter in the fit. The variable $\bm{S}$ obeying such an equation is a Ornstein-Uhlenbeck process, which admits a power spectrum matrix \cite{katsoprinakis_measurement_2007, sinitsyn_theory_2016, mouloudakis_effects_2022} given by:
\begin{equation}
    C(\omega) = (R + i\omega\identity)^{-1} G G^T (R^T -i\omega \identity)^{-1}.
\end{equation}The power spectral density (PSD) of $S_z$ is given by the element $C_{zz}(\omega)$, which yields the result of eq.\,(\ref{equation1}). These three lines are used to fit our experimental data as shown  in Fig.\,\ref{figure1} (b,c).


\begin{figure}
    \centering
    \includegraphics[width=\columnwidth]{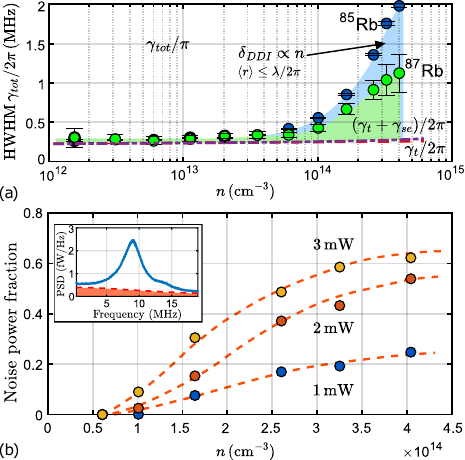}
    \caption{(a) Evolution of the half-width at half maximum (HWHM) of the SN peaks of both isotopes as a function of the Rb vapor density for a probe laser of 2\,mW and a detuning $\Delta/2\pi = 1.5\,\mathrm{GHz}$. The shaded areas mark the part of the linewidth $\delta_{DDI}$ which we attribute to the dipole-dipole coupling. (b) Evolution of the fraction of the total noise power contained in the low-frequency component as a function of the vapor density, for three different probe powers. Dashed lines are guides to the eye.}
    \label{figure2_alt}
\end{figure}

\subsection{Density-dependent broadening of the SN lines}

Using this fitting procedure, more insight into this anomalous lineshape can now be obtained by measuring the half width at half maximum (HWHM) of both spin resonance peaks as the temperature of the cell is varied between $80^{\circ}$C and $180^{\circ}$C. This corresponds to an atomic density spanning  more than two orders of magnitude, from $n \simeq 10^{12} \,\mathrm{at.cm}^{-3}$ to $4 \times 10^{14} \,\mathrm{at.cm}^{-3}$. Figure \ref{figure2_alt}\,(a) shows the resulting trend for both isotopes using the same $\Delta/2\pi = 1.5\,\mathrm{GHz}$ optical detuning. The dashed red line is the value of $\gamma_t/2\pi$, which corresponds to a width limited by the transit of the atoms into the laser beam. The purple dash-dotted line is the sum of the transit and the SE collisional rates $(\gamma_t+\gamma_{se})/2\pi$, using the theoretical expression $\gamma_{se} = K N_{at}\sigma_{se}\bar{v}$, where $\sigma_{se} = 2\times 10^{-14}\,\mathrm{cm}^2$ is the spin-exchange cross section and $K\simeq 0.4$ a factor arising from the nuclear angular momentum of Rubidium \cite{katsoprinakis_measurement_2007}. We recall that they are the two dominant relaxation mechanisms for the one-body dynamics in typical SNS experiments. In our case however, the observed HWHM increases from around $\gamma_t/2\pi = 250\,\mathrm{kHz}$ at low density up to almost 2 MHz at the highest temperature for the $^{85}\mathrm{Rb}$ noise peak. Strikingly, when the density is large enough, the experimental data cannot be explained by the aforementioned relaxation processes, as seen in figure \ref{figure2_alt}\,(a). First, the transit time broadening depends only on the beam size and the atomic velocity, which increases only very weakly with the temperature. Second, the vapor density for which the SE relaxation rate would lead to a broadening of $\sim 2\,\mathrm{MHz}$ is around $5\times10^{16}\,\mathrm{at.cm}^{-3}$. This is two orders of magnitude higher than the actual density used in our experiment, eventually ruling out the possibility that transit and SE broadening only are involved. Finally, it cannot either be explained by the ground-state population excitation rate induced by the probe laser, which is of the order of $ 30\,\mathrm{kHz}$ and independent of the density.
Hence, we introduce  $\delta_{DDI}$, a new contribution to the linewidth defined as:
\begin{equation}
    \frac{\gamma_{tot}}{2\pi}= \frac{\gamma_t + \gamma_{SE}}{2\pi}  + \delta_{DDI}
        \label{equation4}
\end{equation}
and highlighted in shaded areas in Fig.\,\ref{figure2_alt}(a). Let us comment  eq.\,(\ref{equation4}). First, if we cast the critical density of $n= 10^{14}\,\mathrm{at.cm}^{-3}$ above which  $\delta_{DDI}$ becomes significant into an average interatomic distance, we find $\langle r \rangle \simeq 0.55 \,n^{-1/3} \simeq 120\,\mathrm{nm} \simeq \lambda/2\pi $. The fact that this distance is smaller than the wavelength of the light ($780\,\mathrm{nm}$) suggests that resonant dipole-dipole interaction (DDI) induced by the probe light between the atoms in the vapor is responsible for the emergence of $\delta_{DDI}$. We strenghten this claim in the next sections by developing a model that reproduce the variation of $\delta_{DDI}(n, P, \Delta)$ observed experimentally.
Second, while it is always possible to phenomenologically model this broadening by adding a decay rate, it is unclear why such an optical interaction would decrease the ground-state spin coherence lifetime $T_2 = 1/\gamma_{tot}$. The physical meaning of $\delta_{DDI}$ will thus also be clarified by our theoretical model described in section III.

\subsection{Additional low-frequency noise}

Let us now describe the low-frequency tail appearing for atomic densities higher than $n = 10^{14}\,\mathrm{at.cm}^{-3}$. This noise cannot be attributed to shot noise as we measured it independently and subtracted it from the data. This extra component is highlighted in color in Fig.\,\ref{figure1}\,(c), which shows a spectrum obtained for a temperature of $T=175^{\circ}$C ($n = 3\times 10^{14}\,\mathrm{at.cm}^{-3}$) and a probe power of $2\,\mathrm{mW}$. As explained in the previous subsection, we fit this broad line using a Lorentzian function centered on the zero frequency, with amplitude and width being free parameters. 

The precise measurement of its amplitude or its linewidth in a large density range is experimentally difficult as the signal overlaps strongly with the broad SN lines. Hence, we study a more accessible quantity:  the ratio of the noise power contained in the low frequency component over the total spin noise power as a function of the density of the vapor. Figure \ref{figure2_alt}\,(b) shows that this fraction increases with the vapor density. It is not measurable with precision below $n = 10^{14}\,\mathrm{at.cm}^{-3}$, but increases afterwards up to non-negligible values. For a $3\,\mathrm{mW}$ light beam and a density of $n = 4\times 10^{14}\,\mathrm{at.cm}^{-3}$ (180$^{\circ}$C), almost half of the total power is contained in the low-frequency component. Furthermore, the laser power has a large impact on the variance of this noise: the fraction of the total spin noise contained in the low frequency component is three times larger at the highest density for a $3\,\mathrm{mW}$ probe power than for $1\,\mathrm{mW}$. 
As compared to the SN lines described previously, the specificity of this low frequency noise is that it does not exist at low probe power or low density. This strongly suggests that (i) the low-frequency noise is a direct evidence of the dipole-dipole interaction in the vapor, and (ii) it is a spectral feature created by the dynamics of coupled atoms, meaning that it is unlikely to appear in the usual single-particle Bloch equation describing spin noise in non-interacting systems.

\section{Microscopic model of two-body spin fluctuations}

In the previous section, we described the anomalous density dependence of the spin noise spectrum. Such new experimental observations suggest the emergence of a strong coupling regime. In this section, we introduce a simple microscopic two-body model that we further use to validate the claims made in section II.

\begin{figure}
    \centering
    \includegraphics[width=0.9\columnwidth]{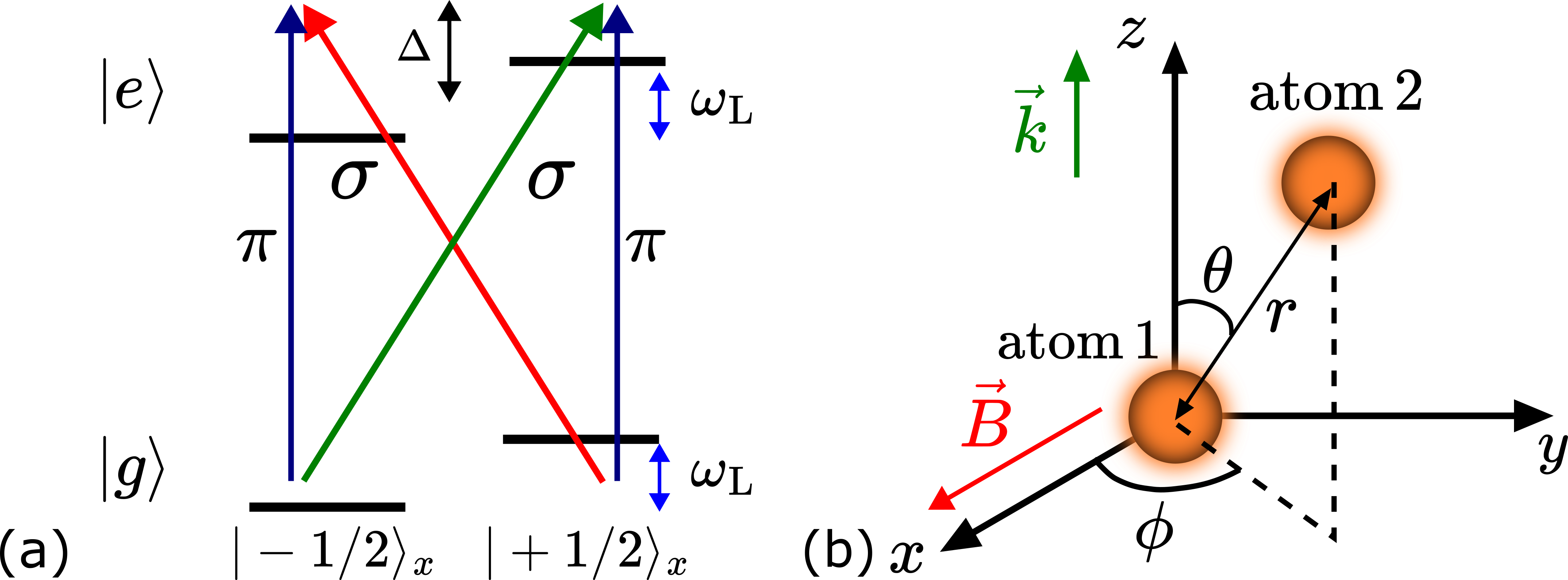}
    \caption{(a) Single-particle energy diagram considered in our model, quantized in the $x$ direction, along the magnetic field. $\pi$ ($\sigma$) denotes light polarized along the $x$ ($y$) axis. The light is detuned by $\Delta$ from the center of the resonance. (b) Schematics and parameters used to describe the spatial arrangement of the two-body problem: one atom is at the origin of the frame and the spherical coordinates give the position of the second atom.}
    \label{figure3}
    \vspace{-0.5cm}
\end{figure}

We propose to describe the spin noise in the ensemble as originating from pair of atoms coupled by resonant dipole-dipole interaction. Since we can neglect the SE collisions rate compared to the transit rate, we make the assumption that inter-hyperfine coherences play no role. We can thus overlook the hyperfine structures of the atoms and consider only one ground-state and one excited state. Moreover, we do not aim at describing higher-order tensorial arrangements of spins \cite{fomin_spin-alignment_2020, liu_spin-noise_2023} but only spin orientation. Consequently, it is reasonable to consider only a $J=1/2$ ground state to simulate the dynamics of spin coherences. The magnetic field is aligned along the $x$ axis and thus splits  the ground-state Zeeman sublevels by an energy $\hbar \omega_L$. Figure \ref{figure3}(a) shows a schematics of this configuration, with a quantization axis in the direction of the magnetic field. Since we want to investigate the action of the probe beam on the optical coupling, we also include a $J'=1/2$ excited state. The light then couples the ground-state Zeeman manifold to the excited one. As the light propagates along the $z$ axis, a linear polarization along the $x$ axis ($\pi$-polarized) couples ground- and excited-states Zeeman sublevels with the same spin projection $m_S$ along the $x$ axis. Orthogonally polarized light ($\sigma$ polarization) couples Zeeman sublevels of opposite values of $m_S$. We recall that in this picture, the spin noise $\braket{S_z(t)}$ arises from fluctuating coherences between the ground-state Zeeman sublevels $\ket{g,-1/2}_x$ and $\ket{g,+1/2}_x$, where $g$ denotes the ground-state.

Let us now consider two atoms with the single-atom energy structure described above. The non-interacting part of the two-body Hamiltonian is thus $\displaystyle{H_0 = \sum_{i=1,2} H_ {SA}^{(i)}}$, with $\displaystyle{H_{SA}^{(i)} = \hbar \omega_L S_x^{(i)} -\mathbf{D}^{(i)}\cdot \mathbf{E}(\bs{r}_i,t)}$, where $\bs{r}_i$ denotes the position of the atom $i$, is the single-atom hamiltonian including Zeeman splitting and light-matter coupling. Here, we choose to work in the long wavelength approximation, i.e. we assume that the atoms are close enough to neglect the dependence on $\bs{r}_i$ of the electric field $\mathbf{E}(\bs{r}_i,t) = \mathbf{E}(t)$. We thus neglect the phase difference in the oscillations of the atomic dipoles. This approximation is justified by our experimental observations: the impact of the DDI on the SN spectra is shown to be relevant only when the interatomic distance is smaller than $\lambda/2\pi $, i.e. when the approximation becomes valid. For large interatomic separation, this approximation breaks down: however, as discussed later in the paper, the radiative coupling has no impact in this case. 

Let us now include the dipole-dipole interaction between the two atoms. This can be done by considering the interaction of both atoms with the vacuum modes of the electromagnetic (EM) field bath \cite{lehmberg_radiation_1970, agarwal_quantum_2012, le_kien_nanofiber-mediated_2017, reitz_cooperative_2022}. Following the standard derivation of the master equation, we trace out the bath degree of freedom to get the following equation of motion, with $\rho$ a two-particle density matrix:
\begin{equation}
     \dot{\rho} = -\dfrac{i}{\hbar} \left[H_0+V_{dd}, \rho\right] + D[\rho] +f,
     \label{ME}
 \end{equation}
where the resonant dipole-dipole interaction Hamiltonian is
\begin{equation}
    V_{dd} = \hbar \sum_{\substack{i\neq j\\\lambda,\lambda'}}\zeta_{\lambda,\lambda'}(r_{ij}) \frac{D_+^{\lambda,i} D_-^{\lambda',j}}{d_0^2}.
    \label{Vdd}
\end{equation} Here, $D_{+}^{\lambda,i}$ (resp. $D_{-}^{\lambda,i}$) is the excitation (resp. de-excitation) part of the dipole operator of atom $i$ and of polarization component $\lambda =\sigma_+, \sigma_-,z$. We denote $d_0$ the dipole element for the $J\rightarrow J'$ transition and $\Gamma_0$ the excited states population relaxation rate. The quantity $\displaystyle{\zeta_{\lambda,\lambda'}(r_{ij}) = -\dfrac{3 \Gamma_0}{4} \, \left(\bs{\varepsilon}_{\lambda}\right)^*\cdot\mathrm{Re}\qty(\rttensor{G}(\mathbf{r}_i, \mathbf{r}_j, \omega_0)) \cdot\bs{\varepsilon}_{\lambda'}}$, where $\displaystyle{\rttensor{G}(\mathbf{r}_i, \mathbf{r}_j, \omega_0) = \qty(\rttensor{\identity}+\dfrac{\bs{\nabla}\bs{\nabla}}{k_0^2} ) \dfrac{e^{i k_0 r_{ij}}}{k_0 r_{ij}}}$ is the electromagnetic Green's dyadic, gives the strength of the radiative coupling between a dipole in $\mathbf{r}_i$  polarized along $\bs{\varepsilon}_{\lambda'}$ and a second one placed in $\mathbf{r}_j$ oscillating in the direction of $\bs{\varepsilon}_{\lambda}$. Here we denoted $k_0 = \omega_0/c$ the light wave vector and $r_{ij} = \vert \mathbf{r}_{i}- \mathbf{r}_{j}\vert$ the interatomic distance. The symmetrical tensor $\rttensor{\zeta}(r_{ij})$ thus strongly depends on the interatomic geometry, and particularly on the distance between two atoms :
\begin{equation}
\centering
    \begin{array}{c c l}
     \rttensor{\zeta}(r_{ij}) &=&  \dfrac{3\Gamma_0}{4}\left(\dfrac{\cos\xi}{\xi^3}+ \dfrac{\sin\xi}{\xi^2}-\dfrac{\cos\xi}{\xi}\right)\identity\\
     &  &  +\dfrac{3\Gamma_0}{4} \left(\dfrac{\cos\xi}{\xi}- \dfrac{3\cos\xi}{\xi^3}-\dfrac{3\sin\xi}{\xi^2}\right) \mathbf{n}\otimes \mathbf{n},\\
    \end{array}
    \label{zeta_definition}
\end{equation}
where $\xi = k_0 r_{ij}$ and $\mathbf{n} = \mathbf{r}_{ij}/r_{ij}$ is the unit vector in the interatomic direction.
The decoherence matrix $D[ \rho ]$ takes the form
\begin{equation}
    \begin{split}
    D[\rho] = &-\gamma_t (\rho-\rho_0) -\sum_{\substack{i,j\\ \lambda, \lambda'}} \dfrac{\gamma_{\lambda,\lambda'}(r_{ij})}{d_0^2}  [D_+^{\lambda,i} D_-^{\lambda',j} \rho\\
    & + \rho D_+^{\lambda,i} D_-^{\lambda',j} - 2D_-^{\lambda',j}\rho D_+^{\lambda,i}],
    \end{split}
\end{equation}
where 
\begin{equation}
    \gamma_{\lambda,\lambda'}(r_{ij}) =  \dfrac{3 \Gamma_0}{4} \, \left(\bs{\varepsilon}_{\lambda}\right)^*\cdot\mathrm{Im}\qty(\rttensor{G}(\bs{r}_i, \bs{r}_j, \omega_0) )\cdot \bs{\varepsilon}_{\lambda'}
\end{equation} is the decoherence rate due to the coupling with the electromagnetic field modes. The first term in the definition of $D[\rho]$ accounts for the relaxations induced by the transit of the atoms towards the state $\rho_0 = \rho_{th}^{(1)}\otimes \rho_{th}^{(2)}$, which is that of two uncorrelated atom at thermal equilibrium. We also include the spin fluctuations by using a noise operator $f$ similar to the one already introduced in refs. \cite{liu_birefringence_2022, liu_spin-noise_2023, delpy_creation_2023}. This Hermitian, traceless operator introduces white-noise fluctuations in the Zeeman population and coherences. This way, we mimic the Poissonian transit of the atoms through the beam, which is the main mechanism of noise in our case. We note that such a "stochastic master equation" has been shown  in \cite{mouloudakis_quantum_2019} to be an equivalent alternative to the quantum trajectories description of spin noise.

\begin{figure}
    \centering
    \includegraphics[width=0.95\columnwidth]{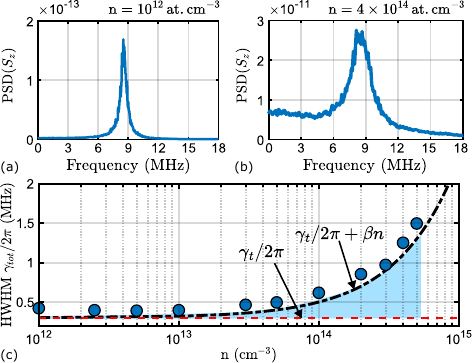}
    \caption{(a,b) Examples of spin noise spectra simulated for densities of $10^{12}\, \mathrm{at.cm}^{-3}$ ($\simeq 75^{\circ}$C) and $4.10^{14}\, \mathrm{at.cm}^{-3}$ ($\simeq 180^{\circ}$C). (c) Simulated evolution of the HWHM of the SN peak as a function of atomic density. Parameters : $\Omega/2\pi = 150\,\mathrm{MHz}$, $\Delta/2\pi = 300\,\mathrm{MHz}$, $\gamma_t = 300\,\mathrm{kHz}$.}
    \label{figure4}
\end{figure}

The equation of motion given by eq.\,(\ref{ME}) is used to simulate the time evolution of the two-atom density matrix. To correctly take into account the action of the light on the atomic populations and coherences, we start from the uncorrelated thermal state $\rho_0$ and let it evolve towards a steady-state density matrix under the action of the magnetic field and the light beam. After it is reached, we add the fluctuations operator in eq.\,(\ref{ME}) and start simulating a set of noisy time traces with a duration $T_0,$ where $T_0$ is chosen much longer than any other typical timescale of the system. For each of them, the ensemble average $\langle S_z\rangle = \mathrm{Tr}(\rho S_z)$ of the spin operator $S_z=\sum_{i=1,2}s_z^{(i)}$ of the two-body system  is computed at each time, along with its Fourier Transform $\hat{F}[S_z] (\omega)$. One then accesses the two-body spin noise PSD $\langle \vert \hat{F}(\omega)^2\vert \rangle$, $\langle ... \rangle $ denoting the average over all simulated samples. It should be mentioned here that we do not average over the Doppler effect experienced by the atoms, and comparison with experimental results should thus be done with parameters that keep the ratio $\Delta/\Gamma$ constant, where $\Gamma$ is the optical coherence decay rate of one atom: this allows to have the same saturation parameter for the experiments and simulations. Nevertheless, since the dipole-dipole interaction depends strongly on the relative positions of the two atoms, we also have to take into account the atomic motion in order to simulate the experimental results. For that purpose, we include random changes in the relative atomic spherical coordinates $(r, \theta, \phi)$ after a time of evolution denoted $\tau_c$. This time can be considered as the time during which the two atoms are close enough for the interaction to exist with a given orientation, before they separate. The interatomic distance $r$ is picked randomly using the nearest-neighbor probability density function and the average atomic density $n$ \cite{ chandrasekhar_stochastic_1943, falvo_double-quantum_2023, yu_long_2019}:
\begin{equation}
    P(r) = 4\pi n r^2 e^{-\frac{4}{3}\pi n r^3}.
\end{equation}

The two angles $\theta, \phi$ follow uniform probability distributions. The spin noise PSD is then computed using the full simulated time-sequence of $\braket{S_z}(t)$, during which $T_0/\tau_c$ changes of conformation occurred. We note here that the correlation time $\tau_c$ is an arbitrary free parameter of the model, set to $\tau_c = 100\,\mathrm{ns}$. A discussion about this choice will be made in the next section.

\section{Results of two-body simulations and discussion}
We now numerically simulate the spin noise spectra associated to two-atoms in interaction and discuss the origin of the broadening $\delta_{DDI}$ as well as its scaling laws with the parameters of the  parameters of the system. We then propose a mechanism for the appearance of the low-frequency noise and briefly comment on its potential to probe the dynamics of atomic binaries.

\subsection{Simulation results and interpretation of the anomalous broadening}
\begin{figure}
    \centering
    \includegraphics[width=\columnwidth]{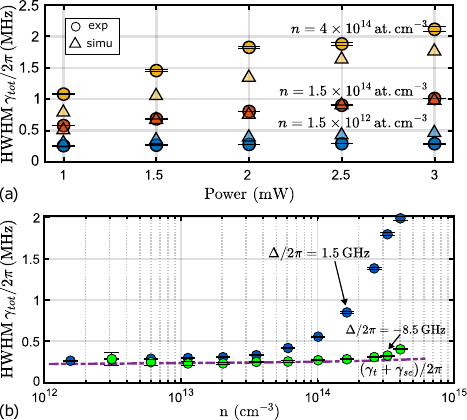}
    \caption{(a) Circles: experimental HWHM of the $^{85}\mathrm{Rb}$ spin noise resonance as a function of laser power, for three different atomic densities. Triangles: Corresponding numerical simulations. (b) Comparison of the experimental broadening of the $^{85}\mathrm{Rb}$ peak induced by the DDI at two different detunings of $\Delta/2\pi = 1.5\,\mathrm{GHz}$ and $\Delta/2\pi = -8.5\,\mathrm{GHz}$}
    \label{figure5_alt}
    \vspace{-0.5cm}
\end{figure}

Figures \ref{figure4}\,(a) and (b) show examples of SN spectra simulated for $n = 10^{12}\, \mathrm{at.cm}^{-3}$ and $10^{14}\, \mathrm{at.cm}^{-3}$ respectively, with a Rabi frequency $\Omega/2\pi = 150\,\mathrm{MHz}$ (matching the 2\,mW probe power used experimentally), a detuning $\Delta/2\pi = 300\,\mathrm{MHz}$, and a Larmor frequency $\omega_L/2\pi = 9\,\mathrm{MHz}$. The shapes of both spectra match very well the experimental results of Figs.\,\ref{figure1}\,(b) and (c). At low densities, the width of the peak equals the transit decoherence rate, and corresponds to the usual single-body dynamics. At high densities the simulations precisely reproduce the broadening of the peak up to the MHz range, as well as the appearance of a broad background noise centered near zero frequency. Figure \ref{figure4}\,(c) shows the evolution of the HWHM of the simulated SN peak with the density. As the latter increases, the average interatomic distance decreases and the resonance broadens. These numerical results are close to the experimental ones highlighted in Fig.\,\ref{figure2_alt}\,(a). In particular, the additional linewidth is found to evolve linearly with the density: $\delta_{DDI}\sim n$, as shown in fig.\,\ref{figure4}\,(c), with a proportionality coefficient $\beta = 2\times 10^{-15}\,\mathrm{MHz.cm}^{3}$. 
The good agreement with the experiment constitutes a good evidence that the changes in the SN spectra observed at high densities are caused by DDI between atoms.

Let us now discuss how such an atomic interaction can actually induce changes in the dynamics of the ground-state coherences and populations. First, the action of the near-resonant laser is to induce small optical coherences \cite{chalupczak_near-resonance_2011, delpy_spin-noise_2023} and therefore a residual excited state population \cite{horn_spin-noise_2011}, which can then give rise to dipole-dipole interaction between an atom in the ground-state and another in the excited state (so-called "resonant" DDI). Consequently, the power of the probe beam should have a huge impact on the interaction strength. Fig.\,\ref{figure5_alt}\,(a) shows the evolution of the HWHM of the $^{85}\mathrm{Rb}$ spin noise as the laser power increases, for temperatures of $T=90^{\circ}$C ($n = 1.5\times10^{12}\,\mathrm{at.cm}^{-3}$), 160$^{\circ}$C ($n = 1.5\times10^{14}\,\mathrm{at.cm}^{-3}$), and 180$^{\circ}$C ($n = 4\times10^{14}\,\mathrm{at.cm}^{-3}$). At low temperatures, the linewidth seems almost independent of the laser power, since the distance between the atoms is too large for the interaction to be relevant. Now, as soon as the average distance becomes shorter than the wavelength: $\langle r\rangle < \lambda/2\pi$ ($T> 150^{\circ}$C i.e. $n> 1\times10^{14}\,\mathrm{at.cm}^{-3}$), the interaction strength becomes non negligible and one can clearly see a net increase in the peak width as the laser drives the dipoles more efficiently. Such observations are well reproduced by our model (see the triangles in Fig.\,\ref{figure5_alt}\,(a)). The discrepancy between experiments and simulations being larger at the highest density may suggest the limit of our two-body model and the need to include three or more particles.

To explain the effects of the DDI on the spin dynamics, we thoroughly study in the supplementary material the spin noise spectra obtained in the static case where the two atoms are fixed in space, and show that the usual SN resonance typical of a one-body evolution now splits into two lines.
The strength of this splitting  depends on the parameter of the two-body geometry, i.e. the triplet $(r, \theta, \phi)$ represented in Fig.\,\ref{figure3}\,(b).  We perform an analytical treatment of the perturbation due to the DDI, and show that this splitting is actually due to a lift of degeneracy in the two-body ground-state energy levels because of the DDI. 
Our calculations demonstrate (see the supplementary material) that $\delta_{DDI}$ scales like
\begin{itemize}
    \item $ 1/r^3 \propto n$ in the limit $\langle r\rangle \ll \lambda/2\pi$;
    \item $ \Omega^2/\Delta^2 \propto P/\Delta^2$ in the weak driving regime.
\end{itemize}

However, the two-body geometry varies with time in the vapor due to the thermal motion. Therefore, over a long measurement time, the changes of the two-body conformation included in the simulations perform an average over different geometries, yielding a single broadened peak centered on the single-body SN line. Hence our model reveals that $\delta_{DDI}$ emerges from the averaging of numerous splitted spectra and thus should be considered as an inhomogeneous broadening of the spin resonance. This solves the apparent contradiction, noted in section II, that the decoherence channels added by the coupling of the atoms to the EM field modes should not be responsible for the relaxation of spin coherences.


The scaling of $\delta_{DDI}$ with the parameters of the experiment thus directly comes from the scaling of the static splitting, being averaged over the distributions of $(r, \theta, \phi)$. Putting all our observations in a single formula, we infer the following scaling for $\delta_{DDI}$, valid at low probe power $P$:
\begin{equation}
    \delta_{DDI}=\kappa \times\frac{P}{\Delta^2}f(n)\Gamma_0,
\end{equation}
with $\kappa$ a constant and $f(n)$ a function of $n$ that emerges from the average over different two-body geometries that scales with $n$ in the large density limit.
The fact that $\delta_{DDI}$ decreases when using a large optical detuning is also experimentally verified: figure \ref{figure5_alt}\,(b) shows that the broadening of the stronger $^{85}\mathrm{Rb}$ peak measured at $\Delta/2\pi = 1.5\,\mathrm{GHz}$ is around 15 times larger than for a detuning $\Delta/2\pi = -8.5\,\mathrm{GHz}$. However, a more detailed study of this point is made difficult by the presence of several hyperfine transitions that contribute to the SN signal.

\subsection{The low-frequency noise as a probe of a two-atoms dynamics}

\begin{figure}[t]
    \centering
    \includegraphics[width=\columnwidth]{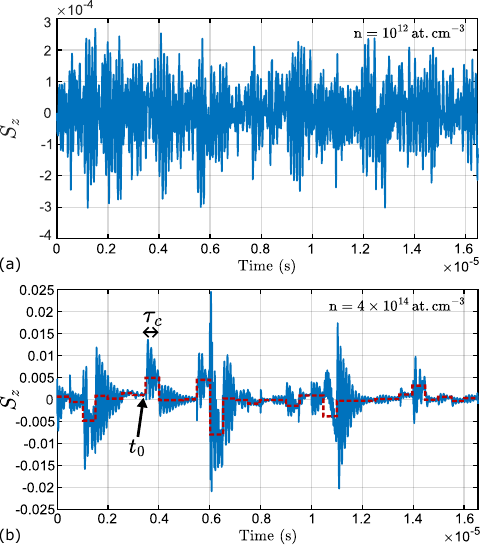}
    \caption{(a) Typical numerical simulations of a time sequence of the two-body spin component $\braket{S_z}(t)$ at a low density of $n=10^{12}\,\mathrm{at.cm}^{-3}$, a 2 mW probe power and a detuning $\Delta/2\pi = 300\,\mathrm{MHz}$. (b) Simulations for a density $n=4\times 10^{14}\,\mathrm{at.cm}^{-3}$. The step-like red dotted line represents the fluctuations of the steady-state $\braket{S_z}^{(st)}$ around which the spins precess.}
    \label{figure6}
\end{figure}

We now propose a mechanism to explain the second unexpected spectral feature, namely the low-frequency noise. Figure \ref{figure6} shows two time traces of spin noise simulated for  different atomic densities. The plot obtained for $n=10^{12}\,\mathrm{at.cm}^{-3}$ [fig.\,\ref{figure6}\,(a)] shows the usual random noise modulated at the Larmor frequency, as expected from the single particle dynamics. However, the result at $n=4\times 10^{14}\,\mathrm{at.cm}^{-3}$, shown in fig.\,\ref{figure6}\,(b), is drastically different. In our simulations, we randomly vary the two atoms conformation every $\tau_c$ in order to mimic the motion of the atom. This sudden change of conformation forces the system to evolve towards a new and non-zero steady-state represented in red dotted-line in fig.\,\ref{figure6})\,(b) due to dipole-dipole coupling. The fluctuations of the value $\braket{S_z}^{(st)}$ towards which the spin relaxes are thus responsible for the low-frequency noise visible on the experimental spectra in figure \ref{figure1}\,(c) and the simulated spectra of figure \ref{figure4}\,(b). We emphasize here that the value of $\tau_c$, which physically represents the lifetime of a binary, has been enlarged to $500\,\mathrm{ns}$ in fig.\,\ref{figure6} to better illustrate the mechanism. In the rest of the paper, we use the value $\tau_c = 100\,\mathrm{ns}$ to match the experimental linewidth of the low-frequency component, which is around $15\,\mathrm{MHz}$.

It is noteworthy that this description makes the low frequency noise a true signature of a two-body dynamics, in the sense that this extra noise can only be ascribed to the dynamics of a pair of atoms and cannot be accounted for by \textit{ad hoc} modification of a single-body equation of motion. Indeed, contrary to the line broadening which could look like an effective reduction of the coherence time $T_2^*$, this noise can simply not be reproduced by the stochastic Bloch equation (\ref{equation2}). This again proves that the spin dynamics in the high-density regime is purely collective and cannot be attributed to a set of identical, independent spins.

However, our model does not allow for a deeper analysis of the binary dynamics for two main reasons. First, the implementation of the atomic motion in the master equation is a highly complex undertaking. Hence, the finite lifetime of the binary is introduced in a simplistic manner as a free parameter that we tune to reproduce the experimental results. Second, our model is not quantitative (it cannot capture properly the amplitude of the noise spectra) and we have to restrict ourselves to the description of the shape of the spectra. Nevertheless, we emphasize here a number of interesting perspectives for the future. Our results suggest that this spectral component reflects the temporal dynamics of a two-atom system, which depends only on the nature and the range of the interaction. Hence we firmly believe that an in-depth study of this noise can reveal more information beyond the mean-field approximation. Investigating the scaling of the low-frequency noise amplitude and width with the density and laser power is further required to help clarifying non-ambiguously its origin. More importantly, as this noise seems to emerge from the correlated dynamics of two particle, we expect possible signatures of atomic correlations in the scaling of the spin noise variance with the atomic density. Nevertheless, the search for a possible departure of the variance from the scaling $\langle S_z^2\rangle\propto n$, valid for an uncorrelated medium, requires (i) a high-bandwidth detection because of the exceptionally large linewidths of the resonances and (ii) a more quantitative model for the two-atom dynamics, which is left for future work.

\section{Conclusion} 

We have experimentally and theoretically investigated anomalous spin noise spectra arising in a dense alkali vapor. Using a mm-thick vapor cell, we studied the Faraday rotation noise imprinted on a laser tuned near the $\mathrm{D_2}$ line of natural Rubidium by the spin fluctuations in the vapor. We took advantage of the low optical depths allowed by such a thin cell to probe high atomic densities and induce strong optical interactions within the vapor. In this regime, we have shown for the first time drastic changes in the spin noise spectra  as compared to the lower densities. We have exhibited two spectral features that are the hallmark of the resonant dipole-dipole coupling, namely a density-dependent broadening of the spin noise lines and a broad noise component centered on low frequencies.

Furthermore, we have introduced a microscopic model to perform numerical simulations and discuss the experimental results. By considering a system of two atoms coupled by resonant dipole-dipole interaction, we have reproduced the unusual experimental lineshapes. Using this model as well as analytical calculations, we have interpreted the unusually large linewidths as an inhomogeneous broadening of the spin precession frequencies due to the dipole-dipole interaction and the atomic motion in the cell. Finally, we have proposed a simple physical picture to explain the emergence of the unexpected low-frequency noise, based on the existence of non-zero spin steady-states for the binaries formed by the dipole-dipole interaction. Eventually, our model suggests that both features are the signature of a genuine two-body dynamics, and cannot be described by the usual mean-field description based on Bloch equations.

This study opens the way to the characterization of many-body spin noise in strongly-correlated ensemble using SNS. Studying a rather simple system such as a dense vapor allowed us to access to collective effects beyond the single-particle model. We believe that a proper description of the low-frequency noise could give further insight on the possibility or not to detect genuine atomic correlation in our experiment. Further researches include a refinement of our model to propose a quantitative analysis of the noise spectra. The inclusion of three or more particles in our simulation is also an exciting prospect, and may already be relevant in the regime presented in this paper. Finally, further work includes the search for possible higher-order spin correlators, which could help revealing the quantum or classical nature of the anomalous spin noise described in this work.

\acknowledgments
The authors are happy to thank Antoine Browaeys, Thomas F. Cutler, Ifan G. Hughes, and Charles S. Adams, for their help with thin cells. We also benefited from the help of Yassir Amazouz and Garvit Bansal for their early interest in the project,  from S\'ebastien Rousselot for technical assistance, and Thierry Jolic\oe ur for some help with simulations. The authors acknowledge funding by the Labex PALM.

\section*{Disclosure}
The authors declare no conflicts of interest.
\section*{Data availability} Data underlying the results presented in this paper are not publicly available at this time but may be obtained from the authors upon reasonable request.


\end{document}


\title{SUPPLEMENTARY MATERIAL: Anomalous high-density spin noise in a strongly interacting atomic vapor}

\author{J. Delpy}
\affiliation{Universit\'e Paris-Saclay, CNRS, Ecole Normale Sup\'erieure Paris-Saclay, CentraleSup\'elec, LuMIn, Orsay, France}
\author{N. Fayard}
\affiliation{Universit\'e Paris-Saclay, CNRS, Ecole Normale Sup\'erieure Paris-Saclay, CentraleSup\'elec, LuMIn, Orsay, France}
\author{F. Bretenaker}
\affiliation{Universit\'e Paris-Saclay, CNRS, Ecole Normale Sup\'erieure Paris-Saclay, CentraleSup\'elec, LuMIn, Orsay, France}
\author{F. Goldfarb}
\affiliation{Universit\'e Paris-Saclay, CNRS, Ecole Normale Sup\'erieure Paris-Saclay, CentraleSup\'elec, LuMIn, Orsay, France}

\maketitle

This supplemental gives more detail about the numerical results of two-body spin noise discussed in the main paper. The first section shows how the dipole-dipole interaction splits the spin noise spectrum in two lines in the case of two atoms fixed in the lab frame. The second section proposes an analytical perturbative treatment of the coupling and discusses how this splitting occurs due to the changes in the ground state energy levels of the binary. 

\section{Simulations of a two-body spin noise spectra in the static configuration}

\subsection{Splitting of the spin noise resonance due to the DDI}

\begin{figure}
    \centering
    \includegraphics[width=\columnwidth]{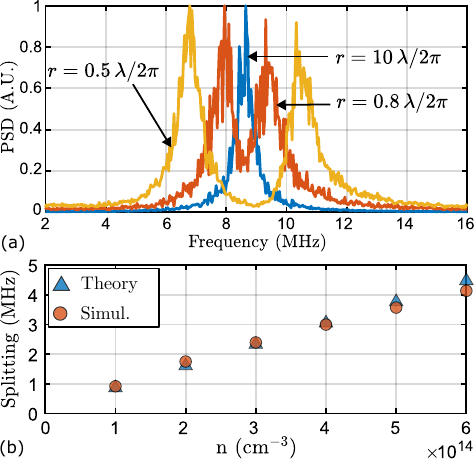}
    \caption{(a) Examples of three spin noise spectra obtained for a fixed geometry with three average interatomic distance $r$ and fixed angles $\theta=0, \,\phi=0$. (b) Orange dots: Numerical calculations of the splitting between the peaks as a function of the density $N = (0.55/r)^3$. Blue triangles: theoretical expression using the calculations of section II.}
    \label{figure1}
\end{figure}

 We here consider the ideal case where the atoms are fixed in the lab frame. The geometry and the definition of the coordinates are shown in figure 3 of the main article. We chose arbitrary spherical angles $\theta$ and $\phi$ between the atoms, as well as a fixed interatomic distance $r$. The two-body problem is fully characterized by the interatomic vector $\mathbf{r}_{ij}$. This in turn determines the Green's tensor $\rttensor{G}(\mathbf{r}_i, \mathbf{r}_j, \omega_0)$ as well as the interaction Hamiltonian and decoherence operators \cite{lehmberg_radiation_1970, agarwal_quantum_2012}. The main result of such a configuration is the following: for large interatomic distances $r>\lambda/2\pi$, the SN spectrum obtained is essentially that of a single spin, while it can dramatically change when $r\leq\lambda/2\pi$. For instance, fig.\,\ref{figure1}\,(a) shows spectra obtained for $\theta =0$ and $\phi =0$, and three interatomic distances. For $r>\lambda/2\pi$, we obtain a single and typical SN resonance. However, for $r\leq \lambda/2\pi$, two lines are clearly visible. This remarkable splitting is plotted as a function of the vapor density from which we infer the average interatomic distance $r$ in figure \ref{figure1}\,(b) (orange dots). As the atomic density increases, it gets larger, eventually reaching several MHz at the highest density. Such a dependency on the average density is discussed in details the next section.  Moreover, there is no effects on the linewidth of the peaks, since the part of the decoherence operator $D[\rho]$ coming from the radiative coupling does not affect the ground-state coherences, but only the optical dipoles. 
 
 However, the interatomic distance $r_{ij}$ is not the only meaningful parameter. Indeed, the angular coordinates $(\theta, \phi)$ rule the intensity of the coupling between the dipoles via the tensor $\mathbf{n} \otimes \mathbf{n}$ appearing in the electromagnetic Green's dyadic. Therefore, the splitting strongly depends on these angles.

\subsection{Influence of the interatomic orientation}

In the previous section, we ran such simulations for fixed angular coordinates $\theta =0,\, \phi = 0$. However, it is clear from the definition of $\rttensor{\zeta}(r_{ij})$, given by eq.\,(8) of the main text, that the angular coordinates also play a crucial role in the strength of the DDI, as we now discuss.
\begin{figure}
    \centering
    \includegraphics[width=\columnwidth]{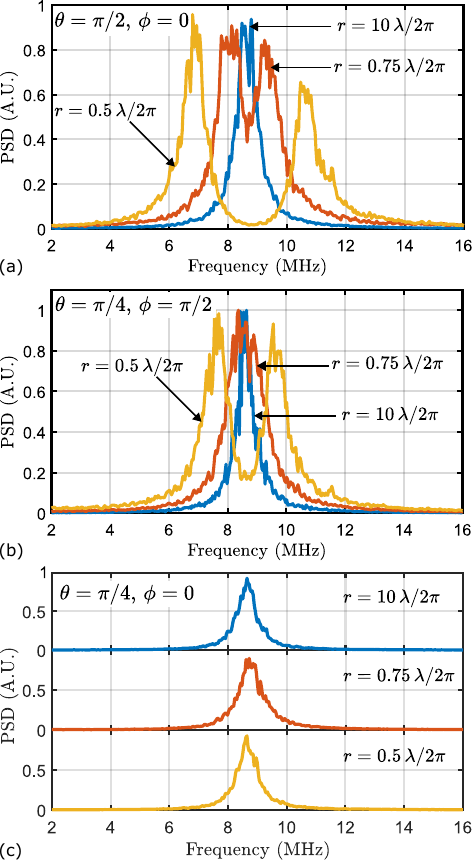}
    \caption{Examples of spin noise spectra obtained
for angular coordinates (a) $\theta = \pi/2$, $\phi = 0$; (b) $\theta = \pi/4$, $\phi = \pi/2$ and $\theta = \pi/4$, $\phi=0$. For each set of angles, the transition between the weak and strong coupling regime is shown using three interatomic distances. In the last case, no significant modification of the spectrum is induced by the DDI.}
    \label{figure2}
\end{figure}

We show in figure \ref{figure2}  a few examples of other sets of angular coordinates $\theta,\,\phi$ to show how they affect the splitting of the SN spectra. One can for instance see in fig.\,\ref{figure2}\,(a)  that the choice of $\theta = \pi/2, \phi = 0$ leads to a strong splitting, which is of the same order of magnitude as the configuration $\theta = 0, \phi=0$. However, it is much reduced when $\theta = \pi/4, \phi = \pi/2$ (fig.\,\ref{figure2}\,(b)) and vanishes completely when $\theta = \pi/4$ and $\phi = 0$ (fig. \,\ref{figure2}\,(c)). 

Understanding this dependence is non trivial without using any numerical simulations. The electromagnetic Green's dyadic $\displaystyle{\rttensor{G}(\mathbf{r}_i, \mathbf{r}_j, \omega_0) = \qty(\rttensor{\identity}+\dfrac{\bs{\nabla}\bs{\nabla}}{k^2} ) \dfrac{e^{i k_0 r_{ij}}}{k_0 r_{ij}}}$ characterizes the electric field created in space by an oscillating dipole $\mathbf{d} (\mathbf{r}_1, t)$ according to:

\begin{equation}
    \mathbf{E}(\mathbf{r},t) = \dfrac{1}{4\pi\varepsilon_0}\rttensor{G}(\mathbf{r}, \mathbf{r}_1, \omega_0)\cdot \mathbf{d}(\mathbf{r}_1,t)
\end{equation}
The electrostatic energy $\Delta E_2 = -\left(\mathbf{d_2}\right)^*\cdot \mathbf{E}(\mathbf{r_2},t)$ acquired by the second dipole $\mathbf{d}_2$ is thus governed by the elements of the tensor $\rttensor{\zeta}(r_{12})$. An explicit expression of the electric field $\mathbf{E}(\mathbf{r},t)$ can be found in electrodynamics and optics textbooks \cite{jackson_classical_2009, Griffiths_2023}. It can be separated in three contribution: the quasi-static (near-field) term, dominating for $kr\ll 1$, the radiated (far-field) term ($kr\gg1$) and finally the inductive term ($kr\simeq 1$). 

The near and far-field terms are well known, so that one can somehow have a view of which coupling $\zeta_{\lambda,\lambda'}$ will be non zero by looking at the characteristic field lines in these regime. Unfortunately, they are not relevant in the study presented here for two reasons: (i) the DDI mediated by the far field is very weak, and the associated perturbation of the two-body ground-state cannot be seen experimentally; (ii) the DDI in the near-field is relevant for atomic densities higher than what we can achieve experimentally. Therefore, the relevant interacting regime in the experimental results presented here is closer to $kr \leq 1$, which does not allow for straightforward representation of the electric field, and thus of the shape and strength of the electromagnetic Green's tensor.

\section{Perturbative treatment of the dipole-dipole coupling}

\begin{figure}
    \centering
    \includegraphics[width=\columnwidth]{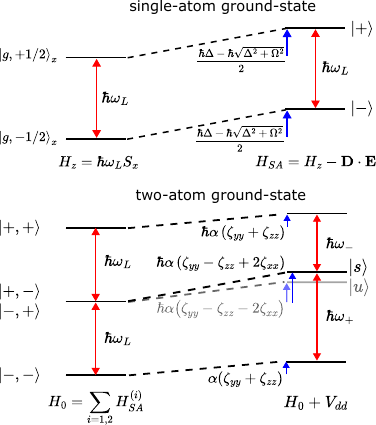}
    \caption{Upper panel: changes in the single-atom ground-state energies due to the light-matter coupling. The perturbed states see their energies impacted by the light shift. Lower panel: changes in the two-atom ground-states induced by the dipole-dipole coupling. The DDI shifts the four state unevenly, resulting in a modification of the spin noise frequencies.}
    \label{figure3}
\end{figure}

The energy level diagram for one particle of the system we consider can be found in figure 4 of the main text. In the absence of probe light, the single-atom eigenvectors are that of the Zeeman hamiltonian, namely $\ket{g, -1/2}_x$ and $\ket{g, +1/2}_x$, with respective energies $-\hbar \omega_L/2$ and $+\hbar \omega_L/2$. This results in the coherences between these two states oscillating at $\omega_L$. The spin noise spectra thus show one peak centered at the Larmor frequency. Let us now add a $\sigma$-polarized probe light, which couples the ground-state sublevel $\ket{g, -1/2}_x$ to the excited state sublevel $\ket{e, +1/2}_x$, and $\ket{g, +1/2}_x$ to $\ket{e, -1/2}_x$. The new low-energy eigenvectors can be computed analytically. In the experimentally relevant case $\omega_L \ll \Delta$, they read
\begin{equation}
\centering
\left\{
    \begin{array}{c c l}
     \ket{-}&=& \cos (\psi/2) \ket{g,-1/2}_x + \sin (\psi/2) \ket{e,+1/2}_x, \\
     \ket{+}&=& \cos(\psi/2) \ket{g,+1/2}_x + \sin (\psi/2) \ket{e,-1/2}_x, \\
    \end{array}
    \right.
    \label{eigenvectors_lower}
\end{equation} where the mixing angle $\psi$ is given by $\tan (\psi/2) = \Omega/(2\Delta)$ with $\Omega$ the Rabi frequency of the light. The Clebsch-Gordan (CG) coefficients do not appear in the mixing angle value, as they are here equal for both transitions. In the weak driving limit ($\Omega\ll \Delta$), the mixing angle tends to zero and we can map directly the state $\ket{-}$ to $\ket{g,-1/2}_x$ and $\ket{+}$ to $\ket{g,+1/2}_x$. These new eigenstates of the one-body hamiltonian $H_{SA}^{(i)}$ are the "light-shifted" states with respective eigenenergies $\displaystyle{\mp\frac{\hbar\omega_L}{2} + \frac{\Delta - \sqrt{\Delta^2+\Omega^2}}{2}}$. This shift in frequency is summarized in figure \ref{figure3} (upper part).
The light shift experienced by $\ket{g, -1/2}_x$ and $\ket{g, +1/2}_x$ are only equal in the approximation $\omega_L \ll \Delta$: in this case, tensorial light shift can be neglected and the states are still separated by an energy $\hbar\omega_L$. The fact that the peak at large $r$ is not perfectly centered on $9\,\mathrm{MHz}$ in the spectra of figs.\ref{figure1} and \ref{figure2} can be explained by this residual tensorial light shift appearing in the simulation. It is worth noting that experiments already demonstrated such lights shifts in spin noise spectra, for instance in metastable helium \cite{delpy_spin-noise_2023}. However, although they were also reported in SNS experiments in alkali atoms \cite{chalupczak_near-resonance_2011}, they are less likely to appear in Rubidium in non resonant conditions because of the small hyperfine splitting of the excited states \cite{happer_light_1970}.

Let us now consider a system of two identical atoms experiencing the influence of both the magnetic field and the light field. We first assume that they are uncoupled, and we move to the two-body Hilbert space. The relevant eigenbasis is thus composed of 16 eigenvectors. Nevertheless, the dynamics of the spin noise emerges from spontaneous coherences arising within the ground-state. The spin noise frequency can therefore be deduced from the structure of the ground-state only, so that we will restrict our analysis to the four lowest energy states. They form the two-body ground-state manifold. Considering the one-atom eigenstates given by eq.(\ref{eigenvectors_lower}), we can construct this ground-state manifold: $\{ \ket{-,-}, \ket{-,+}, \ket{+,-}, \ket{+,+} \}$. However, since the atoms are uncoupled, the two-body SN spectra are similar to that of a single-atom and thus still show one peak at the Larmor frequency.
 
We finally add the dipole-dipole interaction in a perturbative manner. We consider the action of the interaction Hamiltonian $V_{dd}$ on the four uncoupled ground-state eigenvectors of the two-body system.  With the specific choice of angular coordinates ($\theta=0$, $\phi=0$), one can show that the Green's tensor $\rttensor{G}(\bs{r_i}, \bs{r_j}, \omega_0)$ is simply diagonal in the basis $\{x, \,y ,\,z\}$, i.e. the two atoms are coupled via dipoles with similar polarization in this basis. This yields the following expression for $V_{dd}$:

\begin{equation}
\begin{array}{r l}
    V_{dd} =&  \hbar \zeta_{xx}(r_{ij})\left( \dfrac{D_+^{x,1}D_-^{x,2}}{d_0^2} + \dfrac{D_+^{x,2}D_-^{x,1}}{d_0^2} \right) \\
    &+\hbar \zeta_{yy}(r_{ij})\left( \dfrac{D_+^{y,1}D_-^{y,2}}{d_0^2} + \dfrac{D_+^{y,2}D_-^{y,1}}{d_0^2} \right) \\
    &+\hbar \zeta_{zz}(r_{ij})\left( \dfrac{D_+^{z,1}D_-^{z,2}}{d_0^2} + \dfrac{D_+^{z,2}D_-^{z,1}}{d_0^2} \right).
\end{array}
\label{Vdd_xybasis}
\end{equation}

However, since we chose a $\sigma$-polarized light, only $\sigma_+$ and $\sigma_-$ coherences are created, as seen from eq.(\ref{eigenvectors_lower}). It is therefore easier to perform calculations using the dipole operator expressed in the $\{x, \sigma_+, \sigma_-\}$ basis according to the transformation:

\begin{equation}
\left\{
\begin{array}{c c}
    D^{y} = &\dfrac{D^{\sigma_+} + D^{\sigma_-}}{\sqrt{2}},\\
    D^{z} = &\dfrac{D^{\sigma_+} - D^{\sigma_-}}{\sqrt{2}i}.\\
\end{array}
\right.
 \label{D_+-basis}
\end{equation}

These operators write, in the single-atom basis $\{\ket{g,-1/2},\,\ket{g,+1/2},\,\ket{e,-1/2},\,\ket{e,+1/2} \}$:

\begin{equation}
\begin{split}
    D_+^x = \dfrac{d_0}{\sqrt{3}}\begin{pmatrix}
        0 &0 & 0&0\\
        0&0&0&0\\
        -1&0&0&0\\
        0&1&0&0
    \end{pmatrix},\\
    D_+^{\sigma_+} = \sqrt{\dfrac{2}{3}}d_0\begin{pmatrix}
        0&0&0&0\\
        0&0&0&0\\
        0&0&0&0\\
        -1&0&0&0
    \end{pmatrix},\\
    D_+^{\sigma_-} = \sqrt{\dfrac{2}{3}}d_0\begin{pmatrix}
        0&0&0&0\\
        0&0&0&0\\
        0&-1&0&0\\
        0&0&0&0
    \end{pmatrix},
\end{split}
\end{equation}
and obviously $D_-^x = (D_+^x)^\dagger$, $D_-^{\sigma_+} = (D_+^{\sigma_-})^\dagger$ and $D_-^{\sigma_-} = (D_+^{\sigma_+})^\dagger$.

Dropping the dependency of $\zeta_{\lambda\lambda'}$ on $r_{12}$, one can then rewrite eq.(\ref{Vdd_xybasis}) as:

\begin{equation}
\begin{array}{r l}
    V_{dd} =&  \hbar \zeta_{xx} \dfrac{D_+^{x,1}D_-^{x,2}}{d_0^2}   \\
    +& \hbar\left(\dfrac{\zeta_{yy} + \zeta_{zz}}{2d_0^2} \right) \left[D_+^{\sigma_+,1}D_-^{\sigma_-,2} + D_+^{\sigma_-,1}D_-^{\sigma_+,2} \right]\\
    +&  \hbar\left(\dfrac{\zeta_{yy}- \zeta_{zz}}{2d_0^2} \right) \left[ D_+^{\sigma_+,1}D_-^{\sigma_+,2} + D_+^{\sigma_-,1}D_-^{\sigma_-,2} \right]\\
    +&h.c.
\end{array}
\label{Vdd_+-basis}
\end{equation}

\begin{figure}
    \centering
    \includegraphics[width=0.95\columnwidth]{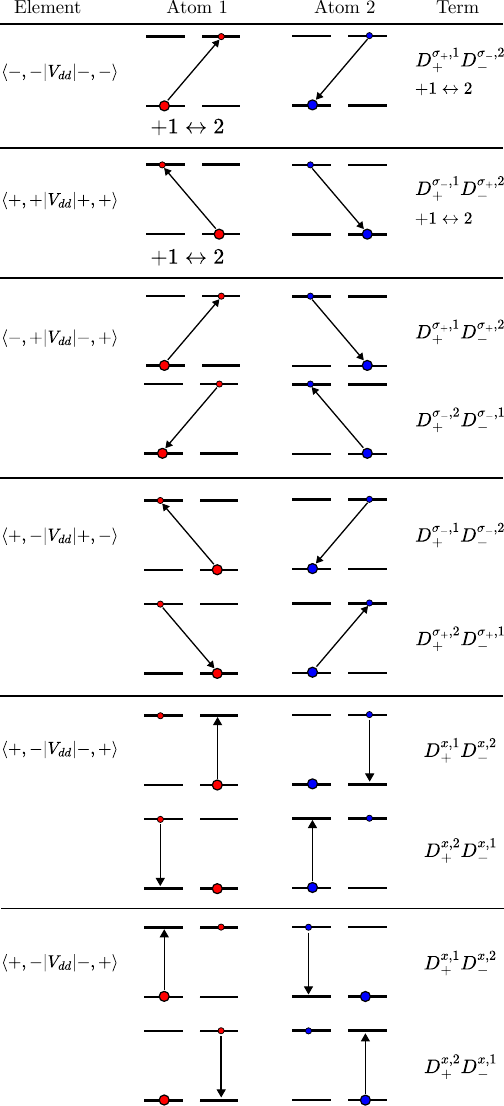}
    \caption{Schematics of the relevant operators in the calculation the first-order energy shifts experienced by the two-body ground-states under the action of the ddi. An upward (downward) arrow designates the action of a $D_+^i\,(D_-^i)$ operator on the atom $i$.}
    \label{figure4}
\end{figure}

Calculating the first-order perturbation induced of the four uncoupled eigenvectors  $\ket{-,-}, \ket{+,-}, \ket{-,+}$ and $\ket{+,+}$ of the two-body lower manifold is now feasible using eqs.\,(\ref{eigenvectors_lower}) and (\ref{Vdd_+-basis}).
The non-vanishing operators in the right-hand side sum of eq.\,(\ref{Vdd_+-basis}) are the ones showing non-zero diagonal matrix element for the four states composing the two-body lower manifold: 
$\{ \ket{-,-}, \ket{-,+}, \ket{+,-}, \ket{+,+} \}$, as well as terms coupling the two degenerate states $\{\ket{-,+}, \ket{+,-}\}$ to one another. These terms are summarized in fig.\,\ref{figure4}, along with the corresponding product of operators.

Consider first the state $\ket{-,-}$: the operators $O$ in the development of $V_{dd}$ with non-zero matrix elements $\braket{-,- \vert O \vert -,-}$ correspond to an excitation of the first atom by a $\sigma_+$-polarized light emitted by the de-excitation of the second atom. Similarly, the state $\ket{+,+}$ is only perturbed at first-order by the same mechanism, except that the emitted light should be $\sigma_-$-polarized. For these mechanisms, the two additional operators obtained by the exchange between atom 1 and 2 must also be considered. This gives the same shift for both states $\ket{-,-}$ and $\ket{+,+}$: 
\begin{equation}
    \braket{-,-|V_{dd}|-,-} =  \braket{+,+|V_{dd}|+,+} = \hbar\alpha(\zeta_{yy} + \zeta_{zz})
\label{energy_shift1_appendix}
\end{equation}
where, letting aside the CG coefficients which are intrinsic to the atomic structure we consider:

\begin{equation}\alpha = \left[ \cos(\psi/2) (\sin \psi/2) \right]^2 = \dfrac{\Omega^2 \Delta^2/4 }{ (\Omega^2/4 + \Delta^2)^2}.
\label{alpha}
\end{equation}

The case of the two other states $\ket{+,-}$ and $\ket{-,+}$ is peculiar due to their initial degeneracy. The corresponding diagonal terms in $V_{dd}$ involve an excitation and de-excitation sequence with radiations of opposite polarizations, as seen in fig.\,\ref{figure4}. Moreover, they are coupled to one another by the absorption and emission process with $x$ polarization, that played no role earlier for the $\ket{-,-}$ and $\ket{+,+}$ states. One can show that in the reduced basis $\{ \ket{-,+},\,\ket{+,-} \}$, $V_{dd}$ reads:

\begin{equation}
    V_{dd} = \hbar\alpha\begin{pmatrix}
        \zeta_{yy}-\zeta_{zz} & 2\zeta_{xx}\\
        2\zeta_{xx} & \zeta_{yy}-\zeta_{zz}.
    \end{pmatrix}
    \label{Vdd_reduced}
\end{equation}

The diagonalization of $V_{dd}$ in this subspace then yields the two symmetrized states $\ket{s} = \dfrac{\ket{+,-} + \ket{-,+}}{\sqrt{2}}$ and $\ket{u} = \dfrac{\ket{+,-} - \ket{-,+}}{\sqrt{2}}$ with respective energy shifts 

\begin{equation}
\left\{
\begin{array}{c c l}
    \braket{s|V_{dd}|s} &=& \hbar\alpha(\zeta_{yy} - \zeta_{zz} + 2\zeta_{xx}),\\
    \braket{u|V_{dd}|u}&=& \hbar\alpha(\zeta_{yy} - \zeta_{zz} - 2\zeta_{xx}),\\
\end{array}
\right.
\label{energy_shift2_appendix}
\end{equation}

Summarizing, one finds that the first-order correction to the two-body ground-state under the action of the resonant DDI is
\begin{equation}
\left\{
\begin{array}{c c l}
    \braket{-,-|V_{dd}|-,-} &=& \hbar\alpha(\zeta_{yy} + \zeta_{zz}),\\
    \braket{+,+|V_{dd}|+,+} &=& \hbar\alpha(\zeta_{yy} + \zeta_{zz}),\\
    \braket{s|V_{dd}|s} &=& \hbar\alpha(\zeta_{yy} - \zeta_{zz} + 2\zeta_{xx}),\\
    \braket{u|V_{dd}|u}&=& \hbar\alpha(\zeta_{yy} - \zeta_{zz} - 2\zeta_{xx}),\\
\end{array}
\right.
\label{energy_shift}
\end{equation}
with $\alpha$ defined in eq.\,(\ref{alpha}). These level shifts due to the DDI are represented in the bottom panel of figure \ref{figure3}. 

The key result of this part is the following: because the ddi lifts the degeneracy between the states $\ket{u}$ and $\ket{s}$, new frequencies appear in the spin noise spectrum. Indeed, the two-atom $z$-axis spin component $\braket{S_z} = \sum_{i=1,2} \braket{s_z^{(i)}}$ precesses at the frequencies given by 
\begin{equation}
    \omega \equiv \frac{E_{\ket{a} }-E_{\ket{b} }}{\hbar},
    \label{new_freq}
\end{equation}
with $\ket{a}$ and $\ket{b}$ two of the two-body ground-states verifying the selection rule $\Delta M_x = \pm1$ \cite{mihaila_quantitative_2006} where $M_x$ denotes an eigenvalue of the two-atom $S_x$ operator.
Fortunately, the four states $\ket{-,-}$, $\ket{s}$, $\ket{u}$ and $\ket{+,+}$ can be directly mapped to the eigenstates of the two operators $\{S^2, S_x \}$ in the weak driving limit. Indeed, since they originate from the addition of two spin 1/2 systems, the two-atom ground-state levels are either spin 1 or spin 0 states:  $\ket{-,-}$, $\ket{+,+}$ and $\ket{s}$ are the spin 1 triplet states, and $\ket{u}$ is the spin 0 singlet state.
The frequencies authorized by eq.\,(\ref{new_freq}) therefore arise from the difference in energy between the states $\ket{-,-}$ and $\ket{s}$ on the one hand, and between the states $\ket{s}$ and $\ket{+,+}$ on the other hand. The spinless state $\ket{u}$ does not participate in the spin noise spectrum of the system. Consequently, taking into account the energy shifts given by eq.\,(\ref{energy_shift}), this yields the frequencies
\begin{equation}
\left\{
\begin{array}{c c}
    \omega_+ =& \omega_L + \delta\\
    \omega_- =& \omega_L - \delta
\end{array}
\right.
\label{new_SN_frequencies}
\end{equation}
where 

\begin{equation}
\delta = \dfrac{1}{2}\dfrac{\Omega^2 \Delta^2}{(\Omega^2/4+\Delta^2)^2}\vert \zeta_{zz} - \zeta_{xx}\vert.
\label{splitting_expression}
\end{equation}

These two frequencies are highlighted in the lower panel of fig.\,\ref{figure3}. This eventually proves how the splitting $2\delta$ visible in figure \ref{figure1}\,(a) is a consequence of the lifting of a degeneracy in the two-body ground-stat. The splitting calculated theoretically using the model presented above are presented in blue triangles in fig.\,\ref{figure1}(b), taking into account the CG coefficients that have been let aside here. The excellent agreement validates our physical picture. Only the last points show a slight discrepancy, suggesting the limit of the perturbative approach above $n = 5\times 10^{14}\, \mathrm{at.cm}^{-3}$.
We emphasize here a couple of remarkable properties of the analytical result of eq.\,(\ref{splitting_expression}). First, the value of the splitting is proportional to a linear combination of elements of the tensor $\rttensor{\zeta}$. Since these elements all scale with the interatomic distance $r$ like a power law $1/r^{\beta}$, this explains why the splitting of fig.\,\ref{figure1} (b) follows the same scaling. This scaling eventually becomes $\delta \propto 1/r^3$ in the limit $r\ll \lambda/2\pi$, as is well known for the dipole-dipole interaction in the near field \cite{agarwal_quantum_2012}. The fact that the splitting represented in fig.\ref{figure1}(b) is not a perfectly linear function of $n$ can be explained by the fact that $r$ is just smaller than $\lambda/2\pi$ but not much smaller in the range of values studied here. We are thus not strictly speaking in the near-field regime. Furthermore, as expected for the coupling between light-induced dipoles, the splitting scales with the Rabi frequency as $\Omega^2$ (i.e. linearly with the laser power) in the weak driving limit. This eventually supports our experimental observations  (see figures 2 and 5(a) of the paper) that the more intense the light beam, the stronger the interaction. 

It has been observed when performing simulations that the polarization of the light matters: when using $\pi$-polarized light, only $\pi$ dipoles are excited among the atoms. One can easily verify that in this case, the diagonal terms of $V_{dd}$ in the uncoupled basis $\{ \ket{-,-}, \, \ket{-,+}\, \ket{+,-},\,\ket{+,+} \}$ arise solely from the $D_+^{x,i}D_-^{x,j}$ part of the hamiltonian and are all equal. In spite of the degeneracy of $\ket{+,-}$ and $\ket{-,+}$ still being lifted by other non-diagonal terms, the resulting splitting is greatly reduced as compared to the earlier case of a $\sigma$-polarized light excitation and barely exceeds the linewidth of each peak.
However, the experimental spin noise spectra bear no such polarization dependence. This can be understood from the limit of our model, where both ground and excited states are composed of two Zeeman sublevels only, creating a strong sensitivity to light polarization. This sensitivity is reduced in atoms having a larger number of sublevels both in ground and excited states, like for instance alkali atoms such as Rubidium. We would thus expect this dependence to be attenuated in the case of a simulation taking into account all hyperfine levels and sublevels with the corresponding CG coefficients.